\documentclass[conference]{IEEEtran}

\ifCLASSINFOpdf
  \usepackage[pdftex]{graphicx}
  \DeclareGraphicsExtensions{.pdf,.jpeg,.png}
\else
\fi

\usepackage[cmex10]{amsmath}
\usepackage{amsfonts}
\usepackage[pdftex]{graphicx}

\begin{document}

\title{Source Coding With Side Information Using List Decoding}

\author{\IEEEauthorblockN{Mortuza Ali}
\IEEEauthorblockA{Dept. Electrical and Electronic Engineering\\
The University of Melbourne, Vic 3010, Australia\\
Email: mortuzaa@unimelb.edu.au}
\and
\IEEEauthorblockN{Margreta Kuijper}
\IEEEauthorblockA{Dept. Electrical and Electronic Engineering\\
The University of Melbourne, Vic 3010, Australia\\
Email: mkuijper@unimelb.edu.au}}

\maketitle

\begin{abstract}
%\boldmath
The problem of source coding with side information (SCSI) is closely
related to channel coding. Therefore, existing literature focuses on
using the most successful channel codes namely, LDPC codes, turbo
codes, and their variants, to solve this problem assuming classical
unique decoding of the underlying channel code. In this paper, in
contrast to classical decoding, we have taken the list decoding
approach. We show that syndrome source coding using list decoding can
achieve the theoretical limit. We argue that, as opposed to channel
coding, the correct sequence from the list produced by the list
decoder can effectively be recovered in case of SCSI, since we are
dealing with a virtual noisy channel rather than a real noisy channel.
%%MK replaced text ``which is a source coding problem.''
Finally, we present a guideline for designing constructive SCSI schemes using Reed Solomon code, BCH code, and Reed-Muller code, which are the known list-decodable codes.
\end{abstract}

\IEEEpeerreviewmaketitle

\section{Introduction}
Recently, in the context of sensor network and mobile multimedia applications~\cite{RamMagaDSC, LiveMagaDSC, GirodProcDVC}, distributed source coding has gained significant attention from the research community. The information theoretic limit for independent encoding of correlated sources has been established by Slepian and Wolf in~\cite{SlepianWolf}. According to the Slepian-Wolf theorem, independent encoding of correlated sources with joint decoding can be as efficient as joint encoding and decoding. More specifically, in the compression of two correlated sources $\{ X_i \}$ and $\{ Y_i \}$ the rates achievable with independent encoding but joint decoding are bounded by $R_X \geq H(X|Y), R_Y \geq H(Y|X), R_X + R_Y \geq H(X,Y)$. In this paper, we focus on the asymmetric approach where $\{ Y_i\}$ is encoded at a rate $H(Y)$ in the conventional way and $\{ X_i\}$ is encoded at a rate $H(X|Y)$ assuming that $\{ Y_i\}$ is available at the decoder. The asymmetric approach is also known as {\em source coding with side information}~(SCSI) in the literature.

The essential idea of distributed source coding is {\em binning}~\cite{CovThoEleInfo5, CsiszarKorner}.  Consider the encoding of a source $\{X_i\}$ in the presence of the side information $\{Y_i\}$ available at the decoder. Let $\mathbb{X}$ and $\mathbb{Y}$ are the alphabets of $\{ X_i\}$ and $\{Y_i\}$ respectively. For large enough $n$, with high probability a source sequence  $\mathbf{x} \in \mathbb{X}^n$ belongs to a set of approximately $2^{nH(X|Y)}$ sequences that are jointly typical with the side information sequence $\mathbf{y} \in \mathbb{Y}^n$. Thus if $\mathbf{y}$ were available both at the encoder and decoder, the outcomes from the source $\{X_i \}$ could be encoded using approximately $H(X|Y)$ bits on average with a very small probability of error. In this case, both the encoder and decoder could construct the same set of jointly typical sequences and use the same indexing leading to correct decoding. However, even if $\{ Y_i \}$ is not available at the encoder, it is possible to achieve the same rate of $H(X|Y)$. The idea is to randomly assign each of the source sequences in $\mathbb{X}^n$ to one of the $2^{nR}$ bins, where $R>H(X|Y)$. Given a source sequence $\mathbf{x}$, the encoding operation is to transmit the index of the bin to which $\mathbf{x}$ belongs and the decoding operation is to choose the sequence $\hat{\mathbf{x}}$ from the indexed bin which is jointly typical with the side information sequence $\mathbf{y}$. Since for large enough $n$ with high probability all the sequences that are jointly typical with a given $\mathbf{y}$ will belong to different bins, with high probability $\hat{\mathbf{x}}$ will be equal to $\mathbf{x}$.

%%MK changed parahraph below
It follows from the above that a practical binning algorithm needs
%Although Slepian-Wolf theorem states the limit on achievable
%compression for SCSI, from its non-constructive proof a practical
%algorithm to achieve this limit does not follow immediately. However,
%it hints at a desirable property of binning: we need
to partition the source data space into a minimum number of bins,
ensuring that for any typical side information sequence each of the
bins contains only one jointly typical source sequence. In other
words, in an appropriate measure of distance it should put as many
source sequences as possible in a bin while maximizing the minimum
distance between any pair of sequences in the bin. Thus each of the
bins can play the role of a good channel code. This connection between
binning and channel codes was first indicated
in~\cite{CommLett:RecResult} while interpreting the Slepian-Wolf
coding. Due to this close connection between binning and channel
coding, most of the SCSI schemes proposed in the literature are based
on the most successful channel codes namely LDPC codes, turbo codes,
and their variants. In these schemes, depending on the conditional
entropy $H(X|Y)$, a channel code of a particular rate needs to be
selected. However, there is always a gap between the compression rate
that can be achieved with a specific channel code and the conditional
entropy for which it can yield near lossless compression. For example,
although the turbo code-based scheme in~\cite{TurboMitranGCOM} and
LDPC-based scheme in~\cite{LiverisLDPC} have been designed for
compression rates of $0.67$ and $0.25$, they
%%MK ``achieve'' instead of ``could achieve''
achieve near lossless compression only when $H(X|Y)=0.49$ and $0.20$
bits, respectively. More importantly, for a given conditional entropy,
there is no guideline for choosing the rate of the code to be designed that can ensure near lossless recovery.

In this paper we present a SCSI scheme based on list
decoding. Although list decoding yields a list of codewords, as
opposed to the classical unique decoding, we demonstrate that the
correct codeword can conveniently be extracted from the list in the
case of SCSI. The main advantage of using list decoding is that it can
improve the compression rate significantly as compared to its
classical counterpart.
%%MK changed text below
Moreover, the approach allows for a guideline for the choice of
channel rate and channel code.
%for a given source, this list decoding based SCSI construct can
%provide a clear guideline about of the choice of the channel code of
%an appropriate rate so that near-lossless recovery with a specified
%error rate is ensured.

The organization of the rest of the paper is as follows. We review the technique of syndrome source coding, a compression technique based on channel codes, in Section~\ref{sec:syndrome}. In Section~\ref{sec:proposed-scheme}, we present the notion of list decoding and describe how this can be used to design a SCSI scheme based on the technique of syndrome source coding. The use of existing list-decodable codes in the design of practical constructive codes for SCSI is detailed in Section~\ref{sec:constructive}. Finally we conclude the paper in Section~\ref{sec:conclusion}.

\section{Syndrome Source Coding}
\label{sec:syndrome}
The most challenging problem in the design of a practical binning
scheme is the systematic construction of bins with algebraic
structures so that the bin indexing and typical set decoding can be
performed with reasonable complexity. In this regard, there is a close
connection between binning and channel codes. A channel code induces a
partitioning of the source data space into cosets that can be indexed
by their respective syndromes~\cite{SloaneTheoErroCode}. If the cosets
of a channel code are such that each of them with high probability
contains only one sequence from the typical set then the cosets
effectively act as bins. In this case the index of the bin
%%MK , or equivalently of the coset to which the source sequence belongs,
can be computed as the syndrome of the sequence.

\begin{figure}[!tb]
\centering
\includegraphics[width=8cm]{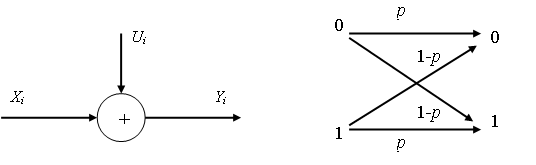}
\caption{An additive noise channel channel can be modelled as a $bSC$. Here, $\{U_i\}$ is an IID Bernolli process with parameter $p$ and thus the crossover probability of the $bSC$ is $p$.}
\label{fig:noise-model}
\end{figure}
Consider encoding a memoryless binary symmetric source $\{X_i\}$ with correlated side information $\{Y_i\}$ available only at the decoder such that
\begin{equation}
\label{eqn:correlation}
Y_i = X_i \oplus U_i
\end{equation}
where $\{U_i\}$ is an IID Bernoulli source with parameter $p< 1/2$. If
$\{Y_i\}$ were present at the encoder as well, it could encode
$\{X_i\}$ at a rate $H(X|Y) = H(U) = -p \lg p - (1-p) \lg
(1-p)$. According to the Slepian-Wolf theorem, $\{X_i\}$ can be compressed at the same rate even if $\{Y_i\}$ is present only at the decoder. This SCSI scenario can be modelled with an additive noise channel where $\{X_i\}$, $\{Y_i\}$, and $\{U_i\}$ correspond to input, output, and noise respectively (see Fig.~\ref{fig:noise-model}). Clearly this additive noise channel is equivalent to a {\em binary symmetric channel} ($bSC$) with a crossover probability $p$ (see Fig.~\ref{eqn:correlation}). This modelling of the correlation between source and side information with a virtual channel allows us to use a channel code for the $bSC$ to design a SCSI scheme~\cite{ZamirNestedLattice} as described below.

The capacity $C_{bSC}$ of the $bSC$ and equivalently of the additive noise channel is
\[
C_{bSC} = 1 - H(U), \quad H(U) = -p \lg p - (1-p) \lg (1-p)
\]
According to the channel coding theorem~\cite{MackayInfoTheoInfAlg4}, there exists an $(n, k)$ linear block code $C$ of rate $k/n = R > (C_{bSC} - \delta)$ such that the error probability $P_e < \epsilon$ for any  $\epsilon > 0$ and $\delta > 0$. Here the error event $\mathbf{x} \neq \hat{\mathbf{x}}$ corresponds to the fact that when the actual noise vector is $\mathbf{u}$, the decoder decides the noise vector to be $\hat{\mathbf{u}}$ and $\mathbf{u} \neq \hat{\mathbf{u}}$. Therefore, $\textnormal{Pr}(\mathbf{x} \neq \hat{\mathbf{x}}) = \textnormal{Pr}(\mathbf{u} \neq \hat{\mathbf{u}})$. Now consider the following scheme of compression of $\{X_i\}$ with the side information $\{ Y_i\}$ available only at the decoder based on this channel code. The encoding operation is to compute the syndrome of a source sequence $\mathbf{x} \in \mathbb{X}^n$ as $\mathbf{s} = \mathbf{H}\mathbf{x}^T$, where $\mathbf{H}$ is the parity check matrix of $C$. If $C_{\mathbf{s}}$ denotes the coset corresponding to the syndrome $\mathbf{s}$, then clearly $\mathbf{x} \in C_{\mathbf{s}}$. The decoding operation is to find the sequence $\mathbf{\hat{x}}$ nearest (in Hamming distance) to the side information sequence $\mathbf{y} \in \mathbb{Y}^n$ in the coset $C_{\mathbf{s}}$. This is equivalent to finding the minimum-weight noise vector $\hat{\mathbf{u}}$ such that $\hat{\mathbf{x}} = \mathbf{y} + \hat{\mathbf{u}}$. Thus the probability of error of the scheme is $\textnormal{Pr}(\mathbf{x} \neq \hat{\mathbf{x}}) = \textnormal{Pr}(\mathbf{u} \neq \hat{\mathbf{u}})$, same as the channel decoding error, which tends to zero as $n \rightarrow \infty$. This coding scheme which can compress $\{X_i\}$ at a rate $(n-k)/n< H(U) + \delta$ with an arbitrarily small probability of error is known as {\em syndrome source coding}~\cite{AnchetaSyndrome}. Clearly, if a channel code of rate $R$ is used for syndrome source coding, the achieved compression rate is $1-R$.

\section{Syndrome Source Coding Using List Decoding}
\label{sec:proposed-scheme}
Clearly, the underlying linear block code $C$ and its associated
 decoding algorithm impact the performance of a syndrome source
 coder. Let $C$ be a $(n, k)$ linear block code over $GF(q)$. In the
 decoding of channel codes, the objective is to find the transmitted
 codeword $\mathbf{c} \in C$, given the received word $\mathbf{r} \in
 GF(q)^n$. The natural decoding approach is to find the codeword which
 has the maximum likelihood of being transmitted given that
 $\mathbf{r}$ has been received. This approach known as {\em maximum
 likelihood decoding}~(MLD) amounts to finding the codeword
 $\hat{\mathbf{c}}$ closest to $\mathbf{r}$ in an appropriate measure
 of distance. However, MLD is known to be NP-complete in
 general~\cite{MDD-NP-complete}. Therefore, {\em bounded distance
 decoding} (BDD) which has greatly reduced complexity is preferred in
 practice that ensures correct decoding only when the number of
errors is upper bounded by
%%MK
% error is bounded below
some error correcting radius $\tau$. Obviously, an unambiguous BDD is possible only if $\tau \leq \lfloor (d_{\min}-1)/2 \rfloor$ where $d_{\min}$ is the minimum distance of the code $C$.

Let us see the implication of BDD for channel coding and in turn for syndrome source coding. For a $bSC$ with crossover probability $p$, the expected Hamming distance between the transmitted codeword $\mathbf{c}$ and the received word $\mathbf{r}$ is $E[d(\mathbf{c}, \mathbf{r})]=np$. Thus for unambiguous decoding, we need a code $C$ with $d_{\min}>2np$. However, the largest rate possible for a binary code of $d_{\min}=2np$ is upper bounded by
%%MK
%is bounded above by
$1-H(2p)$, see~\cite{binaryListGuru}, which is much smaller than the
 capacity of the channel $1 - H(p)$. This in turn implies that the
 compression rate achievable with syndrome source coding that relies
 on BDD is lower bounded
%%MK
% below
by $H(2p)$, which is much bigger than the conditional entropy $H(p)$.

In the above scenario, the main constraint is the requirement of unique decoding which sets the decoding radius to $\lfloor (d_{\min}-1)/2 \rfloor$. One way to circumvent this limitation is to increase the decoding radius beyond $\lfloor (d_{\min}-1)/2 \rfloor$ and allow the decoder to output a list of codewords. This approach will be feasible as long as (i) the list contains only a small number of codewords and (ii) there is an effective way of extracting the correct codeword from the list. The method of decoding beyond $\lfloor (d_{\min}-1)/2 \rfloor$ is known as list decoding in the literature. Let $\mathbb{B}_q(\mathbf{r}, e)$ denote the Hamming sphere of radius $e$ around  a point $\mathbf{r}$ in the space $GF(q)^n$. A code $C$ over $GF(q)$ is said to be $(p, L)$ list-decodable if $|\mathbb{B}_q(\mathbf{r}, np) \cap C| \leq L$. List decoding is considered feasible as long as $L$ grows polynomially with the block length~$n$.

To assess the feasibility of this method, let us look at the
 theoretical limits on list decoding. It has been shown
 in~\cite{algorithmic-ListDecode} that for any integer $L \geq 2$,
 there exists a family of binary linear  $(p, L)$ list-decodable
 channel codes of rate $R \geq 1 - H(p)- 1/L$.
Allowing $L$ to grow, a rate arbitrarily close to the theoretical
 limit $1 - H(p)$ can be achieved. This in turns implies that
%%MK
the corresponding syndrome source coders have compression rate $\leq
%based on list decoding a
% syndrome source coder based on this list-decodable code can achieve a
% compression rate of $
H(p) + 1/L$, which can be made arbitrarily close
 to the conditional entropy $H(p)$ by allowing $L$ to grow.

For non-binary alphabet, the capacity of list decoding is similar to
the binary linear case. It has been shown
in~\cite{algorithmic-ListDecode} that for any alphabet size of $q \geq
2$, list size $L \geq 2$, and $p \in (0, 1-1/q)$, there exists a
family of $(p, L)$ list-decodable $q$-ary channel codes of rate $R
\geq 1 - H_q(p) - 1/L$. Here $H_q(p) = p \log_q(q-1) - p \log_qp -
(1-p) \log_q (1-p)$ is the $q$-ary entropy function. Thus, if the
$q$-ary source $\{X_i\}$ and the side-information $\{Y_i\}$ are correlated in such a way that
\begin{equation}
\label{eqn:q-ary-correlation}
\textnormal{Pr}(X_i \neq Y_i)=p ,
\end{equation}
then $H(X|Y)=H_q(p)$. If a linearity constraint is imposed
%%MK
on the channel code, then the best known limit on list-decoding
capacity~\cite{GuruThesis} turns out to be $R \geq 1 - H_q(p) -
1/\log_q(L+1)$. Consequently, the corresponding $q$-ary syndrome
source coders have a compression rate $\leq
%ing with  $q$-ary linear codes exists that can achieve the entropy
%rate
H_q(p)+1/\log_q(L+1)$. We see that this requires exponentially bigger
list size $L$ than its binary counterpart.

{\em A geometrical interpretation of syndrome source coding using list
  decoding: } Consider the correlation model between the source and
side information as defined
in~(\ref{eqn:q-ary-correlation}). According to the law of large numbers, for large enough $n$, given a side information sequence $\mathbf{y} \in GF(q)^n$, the source sequence $\mathbf{x} \in GF(q)^n$ with high probability will be within a thin shell on the surface of $\mathbb{B}_q(\mathbf{y}, np)$. In fact, the thin shell corresponds to the set of sequences that are jointly typical with $\mathbf{y}$. Now the total number of points in the shell is approximately equal to $|\mathbb{B}_q(\mathbf{y}, np)|$ since for large $n$ almost all the point in $\mathbb{B}_q(\mathbf{y}, np)$ will be in the thin shell. It is known~\cite{algorithmic-ListDecode} that the number of points contained in $\mathbb{B}_q(\mathbf{y}, np)$ is bounded by
\begin{equation}
\label{eqn:spherebound}
|\mathbb{B}_q(\mathbf{y}, np)| \leq q^{nH_q(p)} .
\end{equation}
Now consider syndrome encoding of $\mathbf{x}$ using an $(n, k)$ channel code $C$ over $GF(q)$. Clearly $C$ induces a partitioning of the source data space $GF(q)^n$ into $q^{n-k}$ cosets. Since the points in the shell are uniformly distributed over $GF(q)^n$, for any syndrome $\mathbf{s}$, the number of points in $\mathbb{B}_q(\mathbf{y}, np) \cap C_{\mathbf{s}}$ is approximately $q^{nH_q(p)}/q^{n-k}$. From the list decoding point of view, this syndrome source coding is feasible if $|\mathbb{B}_q(\mathbf{y}, np) \cap C_{\mathbf{s}}|$ is small, which holds only if $R=k/n \leq (1 - H_q(p))$.

{\em Extracting the correct sequence from the list:}  CRC codes are
widely used in practice for error detection. A CRC-$\rho$ code is
defined by a generator polynomial $g(\xi)$ of degree $\rho$ that
assigns $\rho$-bit parity to a sequence. In the setting of
list-decoding based syndrome source coding, the use of a CRC code is
expected to be effective for at least two reasons. Firstly, while in
the context of channel coding, the CRC bits are also subject to
channel noise, this is not the case for syndrome source coding.
%%MK deleted: which is a source coding problem.
In syndrome source coding, we can assume that these CRC bits along
 with the syndrome will be available to the decoder without
 error. Secondly, since the list size $L$ is small
%%MK
(polynomial in $n$), only a few parity bits should be sufficient to
 correctly identify the desired sequence. A CRC-$\rho$ code is
 expected to detect the correct codeword from a list of size~$L$ if
 $\rho \geq \lg L$. Thus a syndrome source coder based on list
 decoding that uses a CRC code to extract the correct sequence has a compression rate of $H_q(p)+1/\log_q(L+1)+ \lg L/n$ which approaches to $H_q(p)$ as $L$ and $n$ grow.

Given a linear $(p, L)$ list-decodable code $C$ over $GF(q)$ of rate
$>1 - H_q(p)$, let us articulate the encoding and decoding operations involved in the syndrome coding of $\mathbf{x}$ in the presence of a side information $\mathbf{y}$ available only at the decoder.

{\bf Encoding: } Let $\mathbf{H}$ be the parity check matrix of the code $C$. Then the syndrome of a source sequence $\mathbf{x}$ can be computed as $\mathbf{s} = \mathbf{H}\mathbf{x}^T$. Let $g(\xi)$ be the generator polynomial of a CRC-$\rho$ code and $x(\xi)$ be the polynomial of degree at most $n-1$ that corresponds to the sequence $\mathbf{x}$. Then the $\rho$-bit CRC of $\mathbf{x}$ corresponds to the polynomial $h(\xi) = x(\xi) \mod g(\xi)$ of degree at most $\rho-1$.

{\bf Decoding:} First we need to list decode $C_{\mathbf{s}}$
 considering the side information $\mathbf{y}$ as the received word.
%%MK
% However,
For this, given any $\mathbf{a} \in C_{\mathbf{s}}$, the list decoding
 algorithm for $C$ can be used
%for this purpose
as follows. Compute
 $\mathbf{y}^\prime = \mathbf{y} - \mathbf{a}$. Using the list
 decoding algorithm for $C$, determine the list $\mathcal{L}$
 consisting of those codewords of $C$ which are within the Hamming
 sphere of radius $np$ around $\mathbf{y}^\prime$. Then adding
 $\mathbf{a}$ to each of the codewords in $\mathcal{L}$, we get the
 list $\mathcal{L}_\mathbf{s}$ consisting of the words from
 $C_\mathbf{s}$ that are at a Hamming distance $\leq np$ from
%%MK which are at most $np$ Hamming distance away from
$\mathbf{y}$. Finally, from $\mathcal{L}_\mathbf{s}$ pick the word
 $\hat{\mathbf{x}}$ such that $\hat{x}(\xi) \mod g(\xi) = h(\xi)$.

{\em Remarks:} The problem of finding any $\mathbf{a} \in C_\mathbf{s}$ amounts to solving the system of linear equations $\mathbf{H}\mathbf{a}^T = \mathbf{s}$. Since there are more unknown than equations, there is at least one nonzero solution to it. Thus, we can find an $\mathbf{a} \in C_\mathbf{s}$ in polynomial time, for example, using Gaussian elimination.

\section{Constructive Code Design}
\label{sec:constructive}
Although the encoding and decoding algorithms presented in the previous section are theoretically sound, there are at least two challenges while designing codes for real-world applications. Firstly, the scheme assumes that $d(\mathbf{x}, \mathbf{y})\approx np$ with high probability, which holds when $n \rightarrow \infty$. However, in the real world we have to operate with finite $n$. Secondly, the scheme also depends on the availability of a $(p, L)$ list-decodable code with an efficient encoder and decoder. To date, efficient list decoding algorithms are known for the families of Reed-Solomon (RS), Bose-Chaudhuri-Hocquenghem (BCH), and Reed-Muller (RM) codes. Associated with each of these codes $C$ is a known list decoding radius $\tau$. Before presenting the main results on these families of codes and their potential in syndrome source coding, in the following we provide a general guideline that can be used to design practical SCSI schemes for the correlation model defined in~(\ref{eqn:q-ary-correlation}).

{\em Block length $n$:}  For better performance it is desirable to
 have $n$ as large as possible.
%%MK added text
However, for large $n$ the computational complexity may become
 problematic.
While RS codes in practical applications are
%%MK
mostly
of length $n=256$ (due to the byte oriented world), it is feasible
%possible
to go up to $n=1024$ with binary BCH and RM codes.

{\em List decoding radius $\tau$:}
%%MK
In theory, for $n$ approaching $\infty$, a
%In ideal case when $n \to \infty$, a code $C$ having a
list decoding radius of $\tau<np+\delta$, where $\delta>0$, is sufficient. In practice, for
fixed $n$ we need to have a list decoding radius of
%%MK changed T into T_\epsilon
$\tau > T_\epsilon$, where $T_\epsilon$ is such that
 $\textnormal{Pr}(d(\mathbf{x}, \mathbf{y})>T_\epsilon)< \epsilon$. It
 can be shown that for large
 $n$ we need to choose $T_\epsilon$ slightly bigger than $np$. Let $r$
 be a random variable representing $d(\mathbf{x}, \mathbf{y})$. Then
 clearly $r$ has a binomial distribution with mean $np$ and variance
 $\sqrt{(np(1-p))}$.
% Thus while the mean $np$ grows as $n$, the standard deviation
% $\sqrt{(np(1-p))}$ grows only as $\sqrt{n}$.
For large $n$, with high probability $r$ will be in the vicinity of $np$. For example, for $n=1000$, $p=0.4$, and $\epsilon=10^{-4}$, the value of $T_\epsilon$ is $459$. If we want to decrease the error probability to $\leq 10^{-5}$, we will need to increase $T_\epsilon$ only by $9$ to $468$.

{\em Code rate $R$:} Since the compression rate achieved with syndrome source coding based on a channel code of rate $R$ is $1-R$, we need to pick a code of largest rate $R$ whose list decoding radius is at least $T_\epsilon$.

{\em CRC code generator $g(\xi)$:} There are a number of standard CRC
codes, see~\cite{Wicker-CRC}. Since the list produced by the list
decoder is guaranteed to be small, a few CRC bits would be enough to
correctly extract the source sequence from the list. In practice, the
CRC-12 is expected to be enough and for $n=1000$ this would incur only $1.2\%$ of redundancy.

In the following we discuss the main list decoding results for the families of RS, BCH, and RM codes and their implication for syndrome source coding.

{\bf RS code:} An $(n, k)$ RS code $C$ over $GF(q)$ is defined by the following parity check matrix $\mathbf{H}$ where $1 \leq k \leq n < q$, $b$ is an integer, and $\alpha$ is an element of $GF(q)$ of multiplicative order $n$.
\[\mathbf{H} = \left(
  \begin{array}{cccc}
    1 & \alpha^b & \cdots & \alpha^{(n-1)b} \\
    1 & \alpha^{b+1} & \cdots & \alpha^{(n-1)(b+1)} \\
    \vdots & \vdots & \vdots & \vdots \\
    1 & \alpha^{b+n-k-1} & \cdots & \alpha^{(n-1)(b+n-k-1)}\\
  \end{array}
\right)\]
It follows from the parity check matrix that $\mathbf{c} \in C$ if and only if $c(\alpha^{b+j})=0$ for all $0 \leq j \leq n-k-1$. RS code can also be defined with the following generator matrix where, $\alpha_1, \alpha_2, \cdots, \alpha_n$ are distinct nonzero element of $GF(q)$
\[\mathbf{G} = \left(
  \begin{array}{cccc}
    1 & 1 & \cdots & 1 \\
    \alpha_1 & \alpha_2 & \cdots & \alpha_n \\
    \alpha_1^2 & \alpha_2^2 & \cdots & \alpha_n^2 \\
    \vdots & \vdots & \vdots & \vdots \\
    \alpha_1^{k-1} & \alpha_2^{k-1} & \cdots & \alpha_n^{k-1} \\
  \end{array}
\right)\]
According to this generator matrix, the codeword corresponding to a message $\mathbf{u}$ can be computed as $\mathbf{c} = (u(\alpha_1), u(\alpha_2), \cdots, u(\alpha_n))$. The family of RS codes are maximum distance seperable (MDS) and thus has $d_{\min} = n-k+1$.

A list decoding algorithm was first discovered for low rate RS codes
by Sudan~\cite{sudanListDecodeRS} and later improved and extended for
all rates by Guruswami and Sudan~\cite{sudanGuruListRS}. For a RS code
of rate $R$, the Guruswami-Sudan algorithm can correct up to $n(1 -
\sqrt{R})$ errors which is clearly beyond half its the minimum distance. Guruswami-Sudan algorithm uses the polynomial representation corresponding to $\mathbf{G}$. Given a received word $\mathbf{r}$, the essential idea of the algorithm is to find the polynomials $u(\xi)$ of degree at most $k$ such that $u(\alpha^j)=r_i$ for at least $\tau$ values of $j\in [0, n-1]$. Recently, Wu~\cite{WuListRS-BCH} has proposed an alternative algorithm that can also achieve the same list decoding radius but with a reduced complexity. Wu's algorithm relies on polynomial representation corresponding to $\mathbf{H}$ and is akin to Berlekamp-Massey algorithm.

{\em Example:} Let us design a SCSI scheme for the correlation model
as defined in~(\ref{eqn:q-ary-correlation}) given that $q=2^8$,
$p=0.3$, $\epsilon = 10^{-4}$ and $n=255$. For these values of $n$ and
$\epsilon$, the value of $T_\epsilon$ turns out to be $T_\epsilon =
105$. Thus we need a code having $\tau > 105$. Using the fact that the
RS code of rate $R$ has $\tau = n(1 - \sqrt{R})$ , we find that the
$(255, 88)$ is the desired RS code. Thus the compression rate
achieved with this scheme is $1 - R = 0.6549$. When $12$ CRC bits are
considered the compression rate increases to $0.702$. In contrast,
unique decoding would require a code with $d_{\min}=2T_\epsilon +1=211$. It is the $(255, 45)$ RS code that has $d_{\min} = 211$. The use of this code for syndrome source coding with unique decoding can only achieve a compression rate of $0.8235$.

{\bf Binary BCH code:} Binary BCH codes can be interpreted as alternate codes of RS codes~\cite{SloaneTheoErroCode}, i.e., if $C_{RS}$ is an RS code over $GF(2^m)$, then $C_{RS} \cap GF(2)^n$ is a BCH code. This interpretation allows the Wu's list decoding algorithm for RS codes to be used for the list decoding of BCH codes. However, Wu~\cite{WuListRS-BCH} has also presented an improved algorithm for list decoding of binary BCH that can achieves a list decoding radius of $\tau = \frac{n}{2}(1 - \sqrt{1-2D})$, where $D=d_{\min}/n$ is the designed relative distance of the BCH code.

{\em Example: } Consider designing a SCSI scheme for the correlation model as defined in~(\ref{eqn:correlation}) for $p=0.2$ and $\epsilon=10^{-4}$. As binary BCH codes of length up to $1023$ can be implemented without any difficulty, we choose $n=1023$. For the given values of $n$, $p$, and $\epsilon$, we calculate $T_\epsilon = 254$. The $(1023, 56)$ BCH code with $D>0.3743$ has $\tau > 382$ and thus can achieve an error probability of $P_e<10^{-4}$ if used for syndrome source coding. The compression rate of this scheme is $0.9453$ which slightly increases to $0.9570$ when $12$ CRC bits are considered. Note that with unique decoding it would need a code of $d_{\min}>508$. The BCH code of designed distance $>508$ is the $(1023, 11)$ code which only achieves a compression rate of $0.9892$.

{\bf RM codes:} For any integers $m$ and $r$ with $0\leq r\leq m$, the $r$-th order binary RM code is an $(n=2^m, k(r, m))$ code having $d_{\min}=2^{m-r}$, where the dimension $k(r, m) = 1 + \binom{m}{1} + \cdots + \binom{m}{r}$. RM codes can be constructed in many ways. Among these, the Boolean function based construction~\cite{ECCCostello} is considered the simplest.

The best known list decoding algorithm is known to be the one by
Gopalan, Klivans, and Zuckerman~\cite{RM-listDecode}. The
Gopalan-Klivans-Zuckerman  algorithm relies on the Boolean function
based construction and achieves a list decoding radius of $\tau = \frac{n}{2}(1 - \sqrt{1-4D})$, where $D$ is the relative distance of the code.

{\em Example:} Consider designing a SCSI scheme for binary correlated
sources using RM codes. Let $p=0.3$ and $\epsilon=10^{-4}$. To operate
in high dimension we choose $m=10$ which corresponds to RM codes of
length $n=1024$. For these values of $n$, $p$, and $\epsilon$, the
required list decoding radius turns out to be $T_\epsilon
=364$. According to the list decoding radius of the
Gopalan-Klivans-Zuckerman algorithm, we need a RM code of $d_{\min} >
235$. The second order RM code is the $(1024, 56)$ code having
$d_{\min}=256$ and thus can ensure the desired probability of
error. The compression rate with this scheme is $0.9453$. When the CRC
bits are included it only increases to $0.9570$. In contrast,
with unique decoding it is not possible to achieve any compression
%%MK
for $p=0.3$ and $\epsilon=10^{-4}$,
since it requires $d_{\min} > 728$ which is only possible for $r=0$.

\section{conclusion}
\label{sec:conclusion}
%%MK I've rewritten the first and last part of the Conclusions.
We applied channel coding ideas on list decoding to the setting
of Distributed Source Coding (DSC). In DSC it is customary to model
the correlation between the source information and the side
information via a virtual channel. In this paper we recognize the
advantage that additional bits can be sent outside of the virtual
channel. We exploit this advantage to accomplish selection of the
correct data from the list obtained by the list decoder. The
additional bits are provided by a CRC code.
%The association of a virtual channel with the correlation between the
%source and side information allows us to use the techniques of channel
%coding for SCSI.
We show that our list decoding-based source coding has a compression rate
that is significantly higher than a classical unique
decoding-based source coder.
%The list decoding approach is well suited for SCSI as
%compared to the channel coding applications. This is due to the fact
%that in case of SCSI, which is a source coding problem, the correct
%sequence can be extracted from the list effectively, for example,
%using CRC codes.
Moreover, the proposed approach has the advantage
that given the conditional entropy, it allows for a clear guideline
for choosing the channel rate and channel code corrsponding to the desired compression rate.
% without resorting to simulations.
% Although syndrome source coding using list decoding achieves better
% compression than that uses classical decoding, there are still
% considerable gaps between the designed compression rates and the
% theoretical limits.
Our future work aims at the design of practical codes through this
approach, in particular the design of efficient
source decoding methods.
%minimizing the gaps
% through improved list decoding.

\section*{Acknowledgment}
This research is supported by the Australian Research Council (ARC).
%under Discovery Project 0987558.

\bibliographystyle{IEEEtran}
\bibliography{IEEEabrv,Thesis}

\end{document}